\documentclass[polish,english]{elsarticle}
\usepackage[T1]{fontenc}
\usepackage[latin2,latin9]{inputenc}
\usepackage{geometry}
\usepackage{color}
\usepackage{float}
\usepackage{amsthm}
\usepackage{amsmath}
\usepackage{graphicx}
\usepackage{babel}
\makeatletter
\numberwithin{equation}{section}
\numberwithin{figure}{section}
\makeatother

\usepackage{babel}
\begin{document}

\title{Analysis of transition between different ringing schemes of the church bell}

\author[rvt]{P. Brzeski}

\author[rvt]{T. Kapitaniak}

\author[rvt,rvt1]{P. Perlikowski\corref{cor1}}

\ead{przemyslaw.perlikowski@p.lodz.pl}

\cortext[cor1]{Corresponding author}

\address[rvt]{Division of Dynamics, Lodz University of Technology, Stefanowskiego
1/15, 90-924 Lodz, Poland}

\address[rvt1]{Department of Civil \& Environmental Engineering, National University of Singapore, 1 Engineering Drive 2, Singapore 117576, Singapore}

\address{{*}przemyslaw.perlikowski@p.lodz.pl}

\begin{abstract}
In this paper we investigate dynamics of church bells, characterize
their most common working regimes and investigate how to obtain them.
To simulate the behavior of the yoke-bell-clapper system we use experimentally
validated hybrid dynamical model developed basing on the detailed
measurements of the biggest bell in the Cathedral Basilica of St Stanislaus
Kostka, Lodz, Poland. We introduce two parameters that describes the
yoke design and the propulsion mechanism and analyze their influence
on the systems' dynamics. We develop two-parameter diagrams that allow
to asses conditions that ensures proper and smooth operation of the
bell. Similar charts can be calculated for any existing or non-existing
bell and used when designing its mounting and propulsion. Moreover,
we propose simple and universal launching procedure that allows to
decrease the time that is needed to reach given attractor. Presented
results are robust and indicate methods to increase the chance that
the instrument will operate properly and reliably regardless of changes
in working conditions. \end{abstract}
\maketitle
\section{Introduction}

Bells are one of the oldest musical instruments which still play an
important cultural role. They were invented in China but today their
sound announces major events all around the world. Depending on the
region bells are mounted in a number of different ways basing on local
customs and tradition. In Europe we can encounter three characteristic
mounting layouts: Central European, English and Spanish. In Central
Europe bells usually tilt on their axis with maximum amplitude of
oscillations below $90$ degrees. In the English system the amplitude
of oscillations is greater and bells perform nearly a complete rotations
in both directions. Conversely, in the Spanish system bells rotate
continuously in the same direction. All these mounting layouts were
developed throughout centuries basing on experience and intuition
of bell-founders and craftsmen. It is common that the bells are cast
using casting mould passed down for ages from father to son and so
forth. Although the design of a bell, its yoke, clapper and a belfry
has been being improved for ages, proper modeling of their dynamics
is still a challenging task. 

The dynamics a yoke-bell-clapper system is extremely complex and difficult
to analyze due to nonlinear characteristic, repetitive impacts and
complicated excitation. Moreover, depending on the mounting layout
different dynamical states can be observed and each type of yoke has
its own specific properties. However, already in 19th century Wilhelm
Veltmann tried to describe mathematically the behavior of the famous
Emperor\textquoteright s bell in the Cologne Cathedral \cite{Veltman1,Veltman2}.
He used simplified equations of motion and explained why the clapper
always remained on the middle axis of the bell instead of striking
its shell. His model was developed basing on the equations of a double
physical pendulum. Heyman and Threlfrall \cite{Lumped_3} use similar
model to estimate inertia forces induced by a swinging bell. The knowledge
of loads induced by ringing bells is crucial during the design and
restoration processes of belfries. That is why there is a number of
works considering dynamic interactions between bell towers and bells
mounted in different manners. Muller \cite{Lumped_7}, Steiner \cite{Lumped_8}
and Schutz \cite{Lumped_9} focus on Central European mounting system
which is also considered in the German DIN standard \cite{Lumped_10}.
We can find similar studies concerning the Spanish system \cite{Ivorra2001,Ivorra20062}
and the English system \cite{Lumped_5,Lumped_6}. Ivorra et. al. present
the comparison between the three mounting layouts and prove that in
the Spanish system forces transmitted to the supporting structure
are significantly lower than in the other two. Results presented in
\cite{Ivorra2009,Lepidi2009} show that in many cases we can improve
bells' working conditions slightly modifying the yoke or its support.

Apart from the studies concerning interactions between bells and their
supports there is a number of publications focusing on the dynamics
of the instrument itself or the clapper to the bell collisions. Klemenc
et. al. contributed with a series of papers \cite{Klemenc2010,Klemenc2012}
devoted to the analysis of the clapper-to-bell impacts. Authors investigate
the consequences of the repetitive hits and compare experimental data
with numerical results obtained from the finite-element model. Presented
results prove that full-scale finite-element model is able to reproduce
the effect of collisions but requires long computational times and
complex, detailed models. Therefore, it would be difficult to use
such models to analyze the dynamics of bells. Because of that, recently
we observe the tendency to use hybrid dynamical models which are much
simpler and give accurate results with less modeling and computational
effort. In \cite{Meneghetti20103363} authors propose lumped parameter
model of the bells mounted in Central European system and prove that
with the model we are able to predict impact acceleration and bell's
period of motion.

In our previous publication \cite{Brzeski_dzwony_1} we present an
improved hybrid dynamical model of the yoke-bell-clapper system. All
parameter values involved in the model have been determined basing
on the measurements of the biggest bell in the Cathedral Basilica
of St Stanislaus Kostka, Lodz, Poland. Proposed model is validated
by comparing the results of numerical simulations with experimental
data. The presented results show that the introduced model is a reliable
predictive tool and can be used for further case studies 

In this paper we describe the most common working regimes of bells
and investigate how the yoke design and the propulsion mechanism influence
their dynamics. To simulate the behavior of the yoke-bell-clapper
system we use the hybrid dynamical model that we present in detail
in our previous publication \cite{Brzeski_dzwony_1}. We characterize
the solutions that can be considered as the proper operation of the
bell and analyze how the geometry of the yoke and the driving motor
output affect the dynamics of the system. We introduce two influencing
parameters and develop diagrams that allow to asses how the bell's
behavior depends on the yoke type and propulsion amplitude. Such plots
describe how presumed working regimes can be obtained and can be beneficial
during the design and/or restoration processes of the bells. In addition,
we investigate the time that is needed to reach given attractor. Presented
results prove that in some cases special launching procedure of the
instrument should be introduced to shorten the time of transient motion.
Finally, we propose simple and universal control of the driving mechanism
that allows to decrease transient time significantly.

The paper is organized as follows. In Section 2 we describe the hybrid
dynamical model of the church bell and introduce parameters that influence
the system's dynamics. In Section 3 we characterize the 7 most common
working regimes and in Section 4 investigate how they can be obtained.
Finally, in Section 5 the conclusions are drawn.

\section{Model of the system\label{sec:Model-of-the}}

The hybrid dynamical model of the yoke-bell-clapper system that we
consider in this paper has been described in detail in our previous
publication \cite{Brzeski_dzwony_1}. To develop the model and determine
its parameters values we have performed detailed measurements of the
existing bell named \textit{``The Heart of Lodz''} (the biggest
bell in the Cathedral Basilica of St Stanislaus Kostka in Lodz). Using
the same bell we have tuned and validated the model by comparing the
results of numerical simulations with the data collected during a
series of experiments. Presented results show that the model is a
reliable predictive tool which can be used to simulate the behavior
of a wide range of yoke-bell-clapper systems. In next subsections
we briefly describe the derivation of the equations of motion and
present the influencing parameters.

\subsection{Geometry of the yoke-bell-clapper system}

The model that we use is build up based on the analogy between freely
swinging bell and the motion of the equivalent double physical pendulum.
The first pendulum has fixed axis of rotation and models the yoke
together with the bell that is mounted on it. The second pendulum
is attached to the first one and imitates the clapper. The photo of
the bell that has been measured to obtain the parameters values is
presented in Fig.\ref{fig:Figure1} (a). In Figs.\ref{fig:Figure1}
(b,c) we show schematic model of the bell indicating the position
of the rotation axes of the bell \inputencoding{latin2}\foreignlanguage{polish}{$o_{1}$}\inputencoding{latin9},
the clapper \inputencoding{latin2}\foreignlanguage{polish}{$o_{2}$}\inputencoding{latin9}
and presenting parameters involved in the model. For simplicity, henceforth,
we use term ``bell'' with respect to the combination of the bell
and it's yoke which we treat as one solid element. 

\begin{center}
\begin{figure}[H]
\begin{centering}
\includegraphics{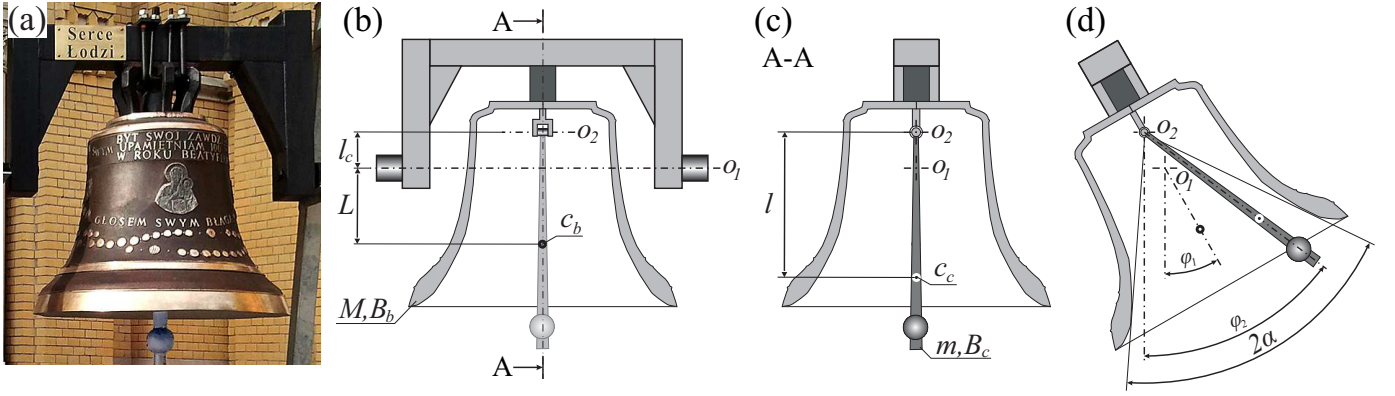}
\par\end{centering}

\protect\caption{\textit{``The Heart of Lodz''} the biggest bell in the Cathedral
Basilica of St Stanislaus Kostka (a) and its schematic model (b,c,d)
along with physical and geometrical quantities involved in the mathematical
model of the system. \label{fig:Figure1}}
\end{figure}

\par\end{center}

The model involves eight physical parameters. Parameter $L$ describes
the distance between the rotation axis of the bell and its center
of gravity (point $C_{b}$), $l$ is the distance between the rotation
axis of the clapper and its center of gravity (point $C_{c}$). The
distance between the bell's and the clapper's axes of rotation is
given by parameter $l_{c}$. The mass of the bell is described by
parameter $M$, while parameter $B_{b}$ characterizes the bell's
moment of inertia referred to its axis of rotation. Similarly, parameter
$m$ describes the mass of the clapper and $B_{c}$ stands for the
clapper's moment of inertia referred to its axis of rotation.

Considered model has two degrees of freedom. In Fig. \ref{fig:Figure1}
(d) we present two generalized coordinates that we use to describe
the state of the system: the angle between the bell's axis and the
downward vertical is given by $\varphi_{1}$ and the angle between
the clapper's axis and downward vertical by $\varphi_{2}$. Parameter
$\alpha$ (see \ref{fig:Figure1} (d)) is used to describe the clapper
to the bell impact condition which is as follows:

\begin{equation}
|\varphi_{1}-\varphi_{2}|=\alpha\label{imp_con}
\end{equation}

Synonymously, collision between the bell and the clapper occurs when
the absolute difference between the bell's and the clapper's angular
displacements is equal to $\alpha$.

\subsection{Equations of motion - modeling of an oscillatory motion of the system
\label{sec:Equations-of-motion}}

In this section we present the mathematical model that we use to simulate
oscillatory motion of the investigated yoke-bell-clapper system. We
use Lagrange equations of the second type and derive two coupled second
order ODEs that describe the motion of the considered system (full
derivation can be found in \cite{Brzeski_dzwony_1}): 

\begin{equation}
\begin{array}{c}
\left(B_{b}+ml_{c}^{2}\right)\ddot{\varphi}_{1}+ml_{c}l\ddot{\varphi}_{2}\cos\left(\varphi_{2}-\varphi_{1}\right)-ml_{c}l\dot{\varphi}_{2}^{2}\sin\left(\varphi_{2}-\varphi_{1}\right)+\left(ML+ml_{c}\right)g\sin\varphi_{1}\\
\\
+D_{b}\dot{\varphi}_{1}-D_{c}\left(\dot{\varphi}_{2}-\dot{\varphi}_{1}\right)=M_{t}(\varphi_{1}),
\end{array}\label{eq1}
\end{equation}

\begin{equation}
B_{c}\ddot{\varphi}_{2}+ml_{c}l\ddot{\varphi}_{1}\cos\left(\varphi_{2}-\varphi_{1}\right)+ml_{c}l\dot{\varphi}_{1}^{2}\sin\left(\varphi_{2}-\varphi_{1}\right)+mgl\sin\varphi_{2}+D_{c}\left(\dot{\varphi}_{2}-\dot{\varphi}_{1}\right)=0.\label{eq2}
\end{equation}
where $g$ stands for gravity and $M_{t}(\varphi_{1})$ describes
the effects of the linear motor propulsion. The motor is active -
and excites the bell - when its deflection from vertical position
is smaller than $\pi/15\:[rad]$ $(12^{o})$. The generalized momentum
generated by the motor $M_{t}(\varphi_{1})$ is given by the piecewise
formula:

\begin{equation}
M_{t}(\varphi_{1})=\begin{cases}
\begin{array}{ccc}
T\,\textrm{sgn}(\dot{\varphi}_{1})\,\cos\left(7.5\varphi_{1}\right), & \: if & |\varphi_{1}|\leq\frac{\pi}{15}\\
\\
0, & \: if & |\varphi_{1}|>\frac{\pi}{15}
\end{array}\end{cases}\label{eq:Mt}
\end{equation}

where $T$ is the maximum achieved torque. Although, the above expression
is not an accurate description of the effects generated by the linear
motor, in \cite{Brzeski_dzwony_1} we prove that it is able to reproduce
the characteristics of the modern bells' propulsions.

There are eleven parameters involved in the mathematical model presented
above. The parameters have the following values: $M=2633\,[kg]$,
$m=57.4\,[kg]$, $B_{b}=1375\,[kgm^{2}]$, $B_{c}=45.15\,[kgm^{2}]$,
$L=0.236\,[m]$, $l=0.739\,[m]$, $l_{c}=-0.1\,[m]$ and $\alpha=30.65^{o}=0.5349\,[rad]$,
$D_{c}=4.539\,[Nms]$, $D_{b}=26.68[Nms]$, $T=229.6[Nm]$. As aforementioned,
all parameters values have been evaluated specifically for the purpose.
For integration of the model described above we use the fourth-order
Runge\textendash Kutta method. ODEs \ref{eq1} and \ref{eq2} together
with the discreet model of impact described in the next Subsection
create a hybrid dynamical system.

\subsection{Modeling of the clapper to the bell impact\label{sec:Modeling-of-the}}

After considering free motion conditions, in this Section we briefly refer to the discreet impact model which is was proposed by Meneghetti and Rossi \cite{Meneghetti20103363} and validated experimentaly in our previous publication \cite{Brzeski_dzwony_1}. When the condition
\ref{imp_con} is fulfilled we stop the integration process. Then,
instead of analyzing the collision course, we restart simulation updating
the initial conditions of equations \ref{eq1} and \ref{eq2} by switching
the bell's and the clapper's angular velocities from the values before
the impact to the ones after the impact. The angular velocities after
the impact are obtained taking into account the energy dissipation
and the conservation of the system's angular momentum that are expressed
by the following formulas:

\begin{equation}
\frac{1}{2}B_{c}\left(\dot{\varphi}_{2,AI}-\dot{\varphi}_{1,AI}\right)^{2}=k\frac{1}{2}B_{c}\left(\dot{\varphi}_{2,BI}-\dot{\varphi}_{1,BI}\right)^{2},\label{ener_dis}
\end{equation}

\begin{equation}
\begin{array}{c}
\left[B_{b}+ml_{c}^{2}+ml_{c}l\cos\left(\varphi_{2}-\varphi_{1}\right)\right]\dot{\varphi}_{1,BI}+\left[B_{c}+ml_{c}l\cos\left(\varphi_{2}-\varphi_{1}\right)\right]\dot{\varphi}_{2,BI}=\\
\\
\left[B_{b}+ml_{c}^{2}+ml_{c}l\cos\left(\varphi_{2}-\varphi_{1}\right)\right]\dot{\varphi}_{1,AI}+\left[B_{c}+ml_{c}l\cos\left(\varphi_{2}-\varphi_{1}\right)\right]\dot{\varphi}_{2,AI}
\end{array}\label{cons_mom}
\end{equation}
where index $AI$ stands for ``after impact'', index $BI$ for ``before
impact'' and parameter $k$ is the coefficient of energy restitution.
In our simulations we assume $k=0.05$ referring to a series of experiments
performed by Rupp et. al. \cite{Lumped_27}. In our previous investigation
\cite{Brzeski_dzwony_1} we have analyzed influence of $k$ on the
response of system and we have proved that dynamics of the system
barely changes for small alterations of parameter $k$ ($\pm20\%$).
Hence, we claim that there is no need to further adjust the value
of $k$ for the considered bell.

\subsection{Influencing parameters}

The mathematical model of the ringing bell described in Section \ref{sec:Model-of-the}
contains 12 parameters ($B_{b}$, $M$, $L$, $D_{b}$, $\alpha$,
$B_{c}$, $m$, $l$, $D_{c}$, $l_{c}$, $k$, $T$) but most of
them are self dependent. Moreover, we have to remember that we consider
a musical instrument, hence we cannot change some of its features
as it could affect the sound it generates. In real applications we
can easily modify the driving motor and the mounting of the bell (by
changing the design of the yoke). Therefore, we try to describe how
changes of the propulsion and the yoke's design influence the system's
dynamics. As a reference, we use values of parameters characteristic
for \textit{``The Heart of Lodz''} and alter them to simulate the
changes in the propulsion or mounting layout.

In our investigation we assume the linear motor propulsion that is
described by piecewise function $M_{t}(\varphi_{1})$ given by the
formula \ref{eq:Mt}. Hence, we can modify the effects generated by
the motor by changing the range of bell's deflection in which the
motor is active (in our case $\left\langle -\pi/15,\,\pi/15\right\rangle $)
or by altering the maximum generated torque $T$. Practically, it
is much easier to modify the maximal momentum which can be realized
either by replacing the motor or by changing the length of the torque
arm. Therefore, we take $T$ as the parameter that describes the driving
motor characteristic and use it as the first controlling parameter.

\begin{figure}
\begin{centering}
\includegraphics{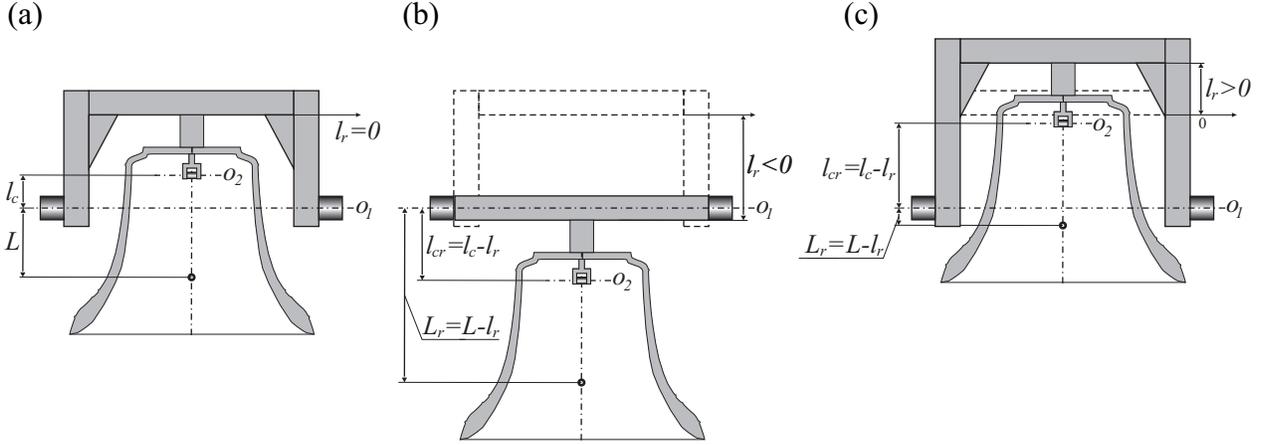}
\par\end{centering}

\protect\caption{The meaning of parameter $l_{r}$: (a) reference yoke design $l_{r}=0$
(based on \textit{``The Heart of Lodz''} yoke), (b) $l_{r}<0$,
(c) $l_{r}>0$.\label{fig:Lr}}
\end{figure}

To describe the modifications of the yoke we introduce new parameter
$l_{r}$ and take it as the second influencing parameter. In Fig.
\ref{fig:Lr} we explain the meaning of $l_{r}$ parameter. 

For the reference yoke (\textit{``The Heart of Lodz''} yoke) we
assume $l_{r}=0$. If we change the yoke design so that the bell's
center of gravity is lowered, then $l{}_{r}<0$ and as the value of
$l_{r}$ we take the distance by which the bell's center of gravity
is shifted with respect to the reference yoke. Similarly, if the bell's
center of gravity is elevated, we assume $l_{r}>0$ and take its displacement
as the value of $l_{r}$. As a maximum considered value of $l_{r}$
we take $0.235\,[m]$ because for $l_{r}=0.236\,[m]$ rotation axis
of the bell goes through its center of mass. 

It is important to point out that changes of $l_{r}$ value affect
other parameters of the model. Hence, whenever we change the value
of $l_{r}$ we have to swap three parameters: $L$ has to be replaced
by $L_{r}$, $l_{c}$ must be replaced by $l_{cr}$, and finally $B_{b}$
by $B_{br}$. Values of $L_{r}$, $l_{cr}$ and $B_{br}$ are calculated
using the following formulas:

\begin{equation}
\begin{array}{c}
L_{r}=L-l_{r}\\
\\
l_{cr}=l_{c}-l_{r}\\
\\
B_{br}=\left(B_{b}-ML^{2}\right)+ML_{r}^{2}
\end{array}\label{r_par}
\end{equation}

\section{Different working regimes of the bell\label{sec:possible behaviours}}

The considered system is not only hybrid and nonlinear but also piecewise
due to the model of driving linear motor. Hence it can exhibit a plethora
of different dynamical phenomena such as periodic, quasi periodic
and chaotic attractors with different number of collisions. Therefore,
a particular bell can behave quite differently depending on the yoke
design and the driving technique. Still, only few types of periodic
attractors can be considered as a proper working regimes and have
practical applications. These regimes are often called ringing schemes
and can be classified in groups that have common characteristics.
The features that are important during the categorization are principally:
the period between the consecutive impacts, the number of collisions
during one period of motion and the course of the collisions. In this
section we point out and describe the seven most commonly meet types
of behavior. We use four types of plots to present considered working
regimes: phase portraits of the bell and the clapper, trajectory projected
in the section of phase space showing relation between the angular
position of the bell and the clapper and time trace of the clapper's
velocity. We use blue lines to present attractors and trajectories;
red lines and arrows are used to present changes of velocity that
are the effects of collisions while green lines are plots of the impact
condition \ref{imp_con} and indicate when the collisions occur.

\subsection{Falling clapper }

We say that the bell works in a ``falling clapper'' manner if the
collisions between the bell and the clapper occur when they perform
an anti-phase motion. This type of behavior is common for bells that
are mounted in the European manner and \textit{``The Heart of Lodz''
}is an example of a bell that works in this regime. In the falling
clapper ringing scheme the amplitude of the clapper's motion is smaller
than the bell's.The clapper's velocity sign changes when collision
occurs. The amount of energy that is transferred during the impact
is relatively large and sufficient to let the bell resound nicely.
Thanks to this feature it is relatively easy to receive nice voicing
of the bells working in falling clapper regime. But, in some cases,
the energy transfer is too abrupt which can lead to crack or damage
of the bell or the clapper. 

\begin{center}
\begin{figure}[h]
\begin{centering}
\includegraphics[scale=0.9]{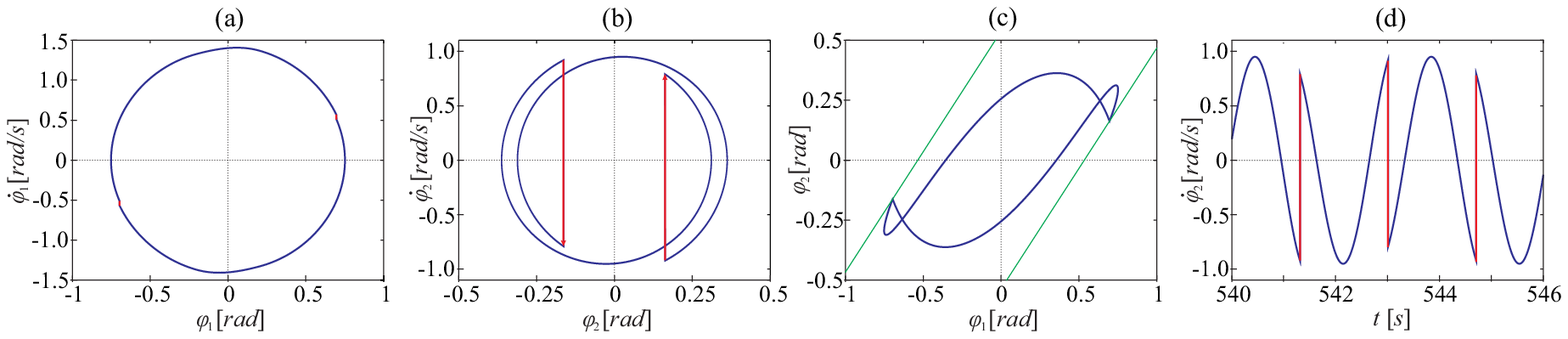}
\par\end{centering}

\protect\caption{Presentation of the symmetrical ``falling clapper'' with $2$ impacts
per period obtained for $T=350\,[Nm]$ and $l_{r}=0.05\,[m]$. Phase
portraits of the bell (a), the clapper (b), section of phase space
with green lines that indicate impact conditions (c) and time trace
of the clapper's velocity (d). Effects of collisions (changes of velocity)
are marked with red lines. \label{fig:Fall_2i} }
\end{figure}

\par\end{center}

We can distinguish a symmetric type of falling clapper with $2$ collisions
per one period of motion and its asymmetric version with $1$ impact
per period. Although the course of the collision is similar in both
cases they are completely different for the listeners as the time
interval between successive impacts is almost doubled in the asymmetric
case. In Figs . \ref{fig:Fall_2i} and \ref{fig:Fall_1i} we present
two types of falling clapper ringing scheme.

\begin{center}
\begin{figure}[h]
\begin{centering}
\includegraphics[scale=0.9]{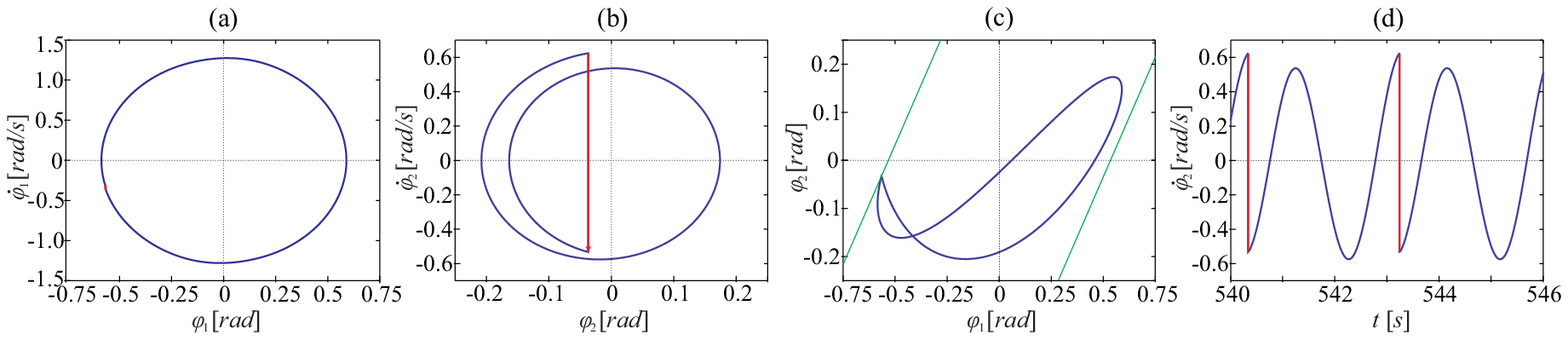}
\par\end{centering}

\protect\caption{Presentation of the asymmetrical ``falling clapper'' with $1$ impact
per period obtained for $T=150\,[Nm]$ and $l_{r}=-0.03\,[m]$. Phase
portraits of the bell (a), the clapper (b), section of phase space
with green lines that indicate impact conditions (c) and time trace
of the clapper's velocity (d). Effects of collisions (changes of velocity)
are marked with red lines.\label{fig:Fall_1i}}
\end{figure}

\par\end{center}

\subsection{Flying clapper}

If the collisions between the bell and the clapper occur when they
perform in-phase motion we say that the bell works in a ``flying
clapper'' manner. In this regime the amplitude of the clapper's motion
is larger than the bell's and the clapper's velocity sign remains
the same after the collisions. The collisions have more gentle characteristic
and the amount of energy that is transferred during the impact is
much smaller than in ``falling clapper'' ringing scheme. Therefore,
sometimes it may be difficult to achieve nice resounding of the bell,
but the risk of the crack or damage is significantly decreased. In
Figs. \ref{fig:Fly_2i} and \ref{fig:Fly_1i} we present two types
of ``flying clapper'' behavior: symmetric attractor with $2$ impacts
per period and asymmetric one with only $1$ impact per period respectively.

\begin{center}
\begin{figure}[h]
\begin{centering}
\includegraphics[scale=0.9]{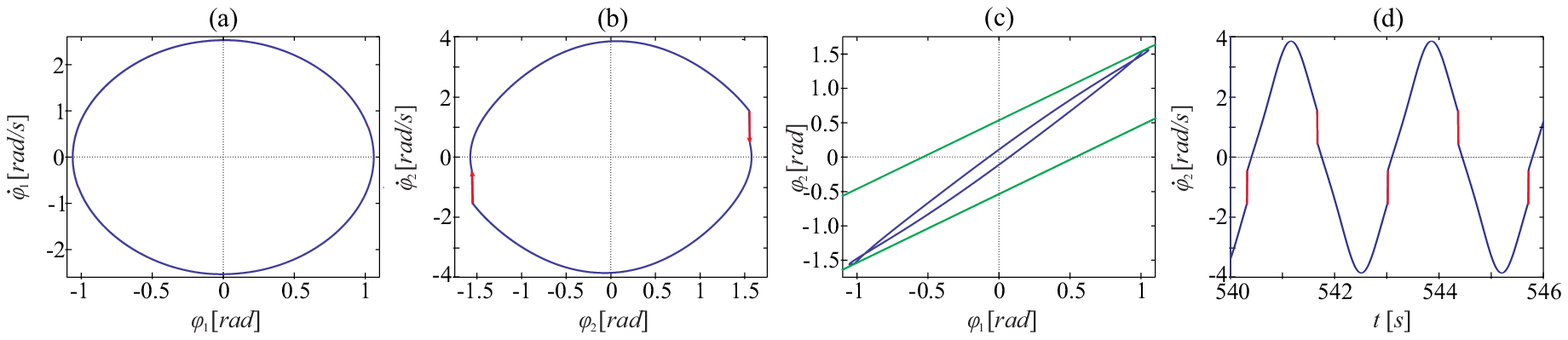} 
\par\end{centering}

\protect\caption{Presentation of the symmetrical ``flying clapper'' with $2$ impacts
per period obtained for $T=450\,[Nm]$ and $l_{r}=-0.91\,[m]$. Phase
portraits of the bell (a), the clapper (b), section of phase space
with green lines that indicate impact conditions (c) and time trace
of the clapper's velocity (d). Effects of collisions (changes of velocity)
are marked with red lines.\label{fig:Fly_2i}}
\end{figure}

\par\end{center}

\begin{center}
\begin{figure}[h]
\begin{centering}
\includegraphics[scale=0.9]{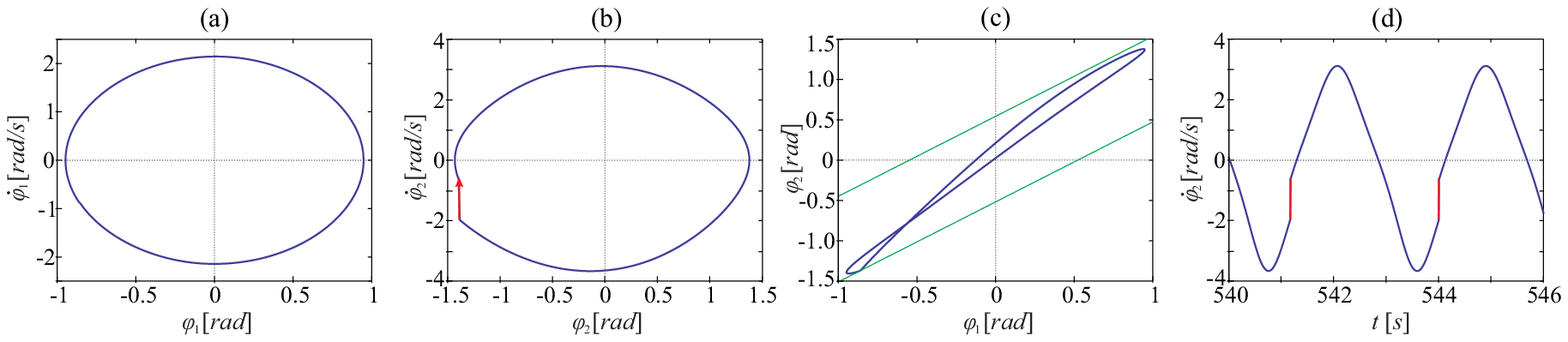}
\par\end{centering}

\protect\caption{Presentation of the asymmetrical ``flying clapper'' with $1$ impact
per period obtained for $T=325\,[Nm]$ and $l_{r}=-1.21\,[m]$. Phase
portraits of the bell (a), the clapper (b), section of phase space
with green lines that indicate impact conditions (c) and time trace
of the clapper's velocity (d). Effects of collisions (changes of velocity)
are marked with red lines.\label{fig:Fly_1i}}
\end{figure}

\par\end{center}

\subsection{Double kiss}

The ``double kiss'' is the name of working regime in which we observe
$4$ impacts per one period of motion. During one period the clapper
hits each side of the bell's shell twice. The first collision on each
side is in the ``falling clapper'' manner while the second impact
has ``flying clapper'' course. This behavior is especially attractive
for the listeners. Still, it is rarely met mainly because it is difficult
to achieve. In the next Section we precisely determine conditions
(ranges of parameters $T$ and $l_{r}$) for which we can observe
this working regime. In Fig. \ref{fig:double_kis} we show the characteristics
of ``double kiss''. 

\begin{center}
\begin{figure}[h]
\begin{centering}
\includegraphics[scale=0.9]{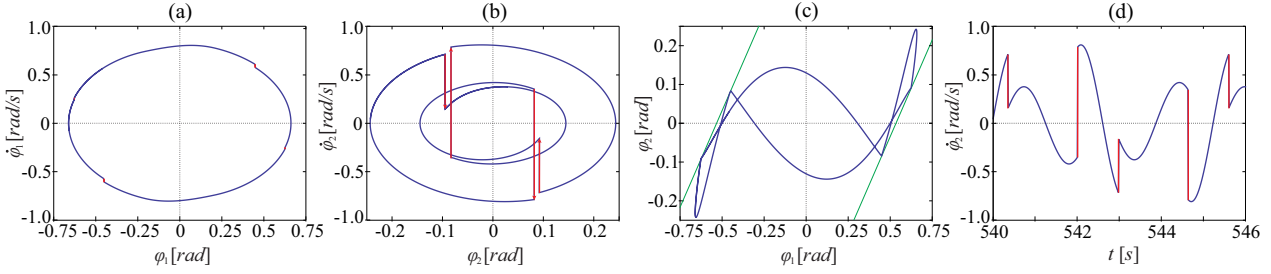}
\par\end{centering}

\protect\caption{Presentation of the ``double kiss'' phenomenon with $4$ impacts
per period. The plots were obtained for $T=175\,[Nm]$ and $l_{r}=0.16\,[m]$.
Phase portraits of the bell (a), the clapper (b), section of phase
space with green lines that indicate impact conditions (c) and time
trace of the clapper's velocity (d). Effects of collisions (changes
of velocity) are marked with red lines.\label{fig:double_kis}}
\end{figure}

\par\end{center}

\subsection{Sticking clapper - sliding dynamics}

``Sticking clapper'' is the name of working regime in which the
clapper and the bell remain in contact for a certain amount of time.
In other words the ``sticking clapper'' refers to the attractors
that contain sliding mode. This working regime is typical for the
bells mounted and operated in the English manner. In the considered
system prior the sliding mode we observe a number of successive impacts
(usually $3$) that have a ``falling clapper'' course. The energy
amount that is transferred between the bell and the clapper decreases
with each subsequent collision. Hence, the sound effects caused by
each hit are different and not all collisions may be noticed by the
listener. Moreover, when the clapper remains in the contact with the
bell's shell it also influences the produced sound. All these features
have to be taken into consideration if we want the bell to sound nicely. 

\begin{center}
\begin{figure}[h]
\begin{centering}
\includegraphics[scale=0.9]{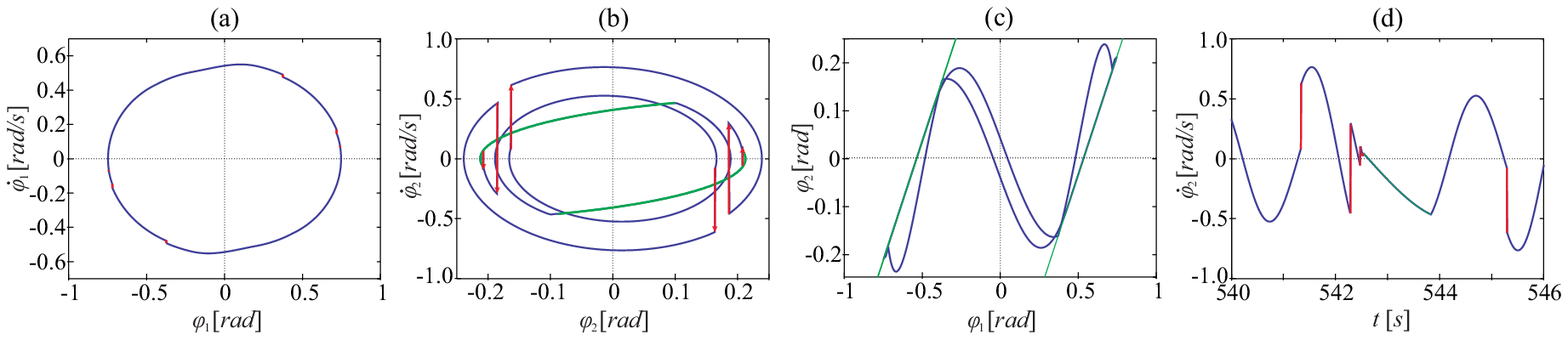}
\par\end{centering}

\protect\caption{Presentation of the ``sticking clapper'' phenomemon. The plots were
obtained for $T=125\,[Nm]$ and $l_{r}=0.2\,[m]$. Phase portraits
of the bell (a), the clapper (b), section of phase space with red
lines that indicate impact conditions (c) and time trace of the clapper's
velocity (d). Effects of collisions (changes of velocity) are marked
with red lines.\label{fig:slide}}
\end{figure}

\par\end{center}

\subsection{No impacting attractors and other possible solutions}

If the forcing amplitude is not sufficient or the yoke is designed
improperly, we can observe stable periodic attractor with no collisions.
In such conditions no sound is produced and the bell can not work
as a musical instrument. Unfortunately, no impacting attractors occur
in wide range of $T$ and $l_{r}$ and even nowadays it is common
that we have to redesign the yoke to make ringing possible. Apart
from the ringing schemes described in this Section we can indicate
other periodic attractors that can be successfully employed. For example
we can observe asymmetric flying clapper behavior with doubled period
and $4$ impacts per period. Such behavior can be easily taken as
a typical falling clapper. Similarly, a quasi-periodic attractor with
almost equal time intervals between the subsequent impacts can sound
almost like a periodic ringing scheme. Moreover, as we proved in our
previous publication \cite{Brzeski_dzwony_1}, the course of successive
collisions can differ a bit. Hence, in practical applications we do
not demand strictly periodic behavior. Instead, we want to receive
similar collisions of a presumed type in a possibly equal time intervals.
Still, when designing the yoke and the propulsion of the bell we should
always look for conditions that ensure periodic behavior as it increases
the probability that the real system will work properly and reliably.

\section{Influence of the yoke design and forcing amplitude on the system's
dynamics}

\subsection{One parameter diagram - transitions between different working regimes}

\begin{figure}[p]
\begin{centering}
\includegraphics{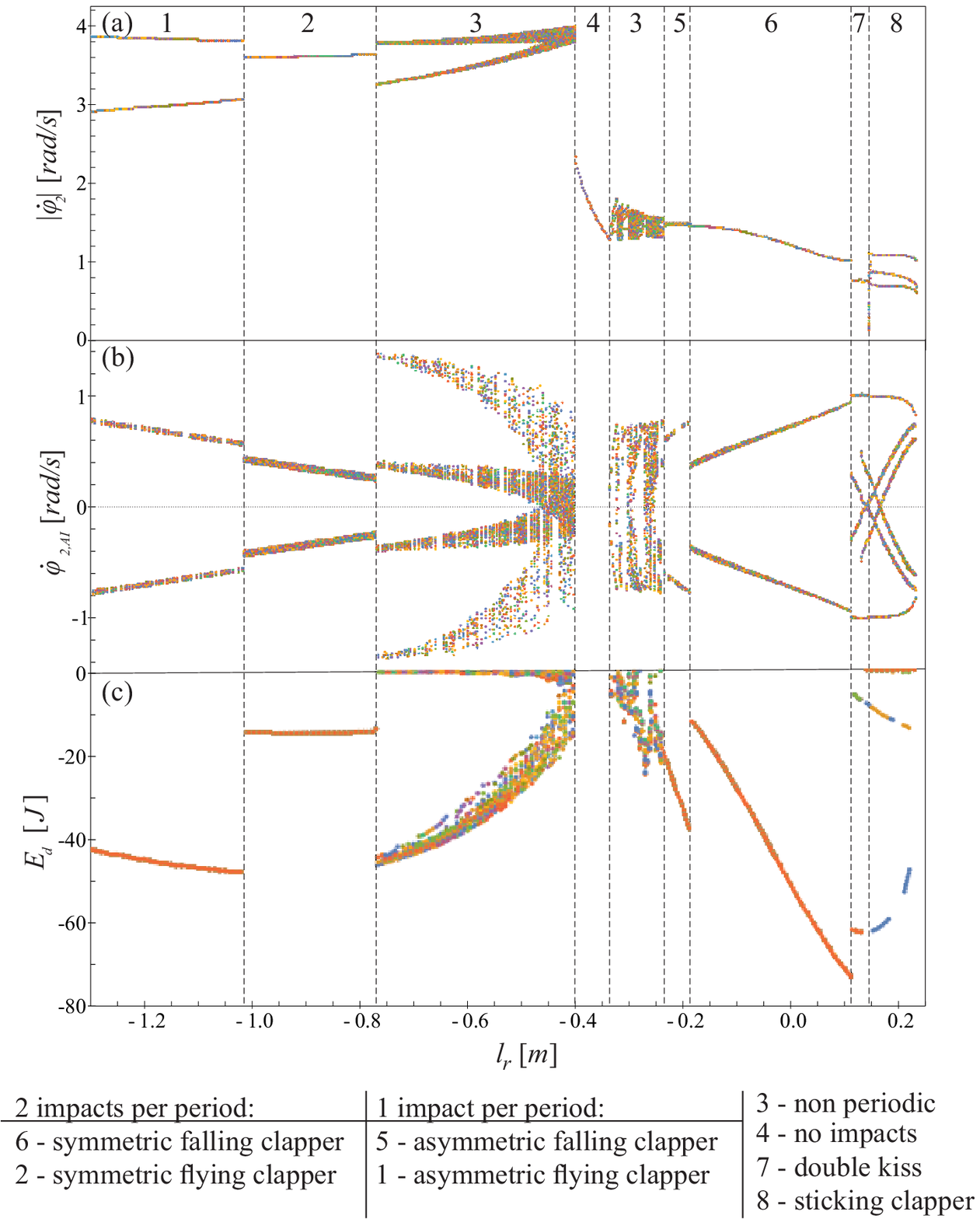}
\par\end{centering}

\protect\caption{One parameter diagram showing the behaviour of the yoke-bell-clapper
system with varying value of $l_{r}$ parameter.  Supblot (a) presents clapper's angular velocity when the clapper passes the downward vertical, (b) shows clapper's velocity just after the impact and subplot (c) presents the energy dissipated at instant of collision. \label{fig:Lr-1a}}
\end{figure}

In Fig. \ref{fig:Lr-1a} we present one parameter diagram that presents
how the system behaves with varying value of $l_{r}$ parameter. We
assume $T=412.5\,[Nm]$ and analyze the dynamics of the system for
$l_{r}\epsilon\left\langle -1.3,\,0.23\right\rangle \,[m]$. Each
time we start integration form zero initial conditions ($\varphi_{1}=0$,
$\varphi_{2}=0$, $\dot{\varphi}_{1}=0$, $\dot{\varphi}_{2}=0$),
hence it is not a bifurcation diagram (the bell and the clapper always
start their motion from hanging down position with zero velocities).
We investigate the behavior of the system after $38$ minutes of preliminary
motion (the analysis of influence of transient time on the system's
dynamics is shown in the end of this Section). To determine in which
regime the system operates, we use two indicators. The first is the
absolute value of the clapper's angular velocity $\left|\dot{\varphi}_{2}\right|$
when the clapper passes the downward vertical $\varphi_{2}=0$. The
second is the clapper's velocity just after the clapper to the bell
impact $\dot{\varphi}_{2,AI}$.  Moreover, we calculate the energy dissipated  at instant of collision $E_{d}$. We record these values during the period of two minutes. Subplots (a) and (b) of Fig. \ref{fig:Lr-1a} correspond to the first and the second indicator respectively while in lower part of the Figure (subplot (d) ) we present the energy dissipation. Analyzing
Fig. \ref{fig:Lr-1a} (a,b) we can asses the periodicity and the symmetry
of the attractor. Moreover, we are able to evaluate the number of
collisions per period of motion. To determine if we observe the ``flying
clapper'' or ``falling clapper'' ringing scheme, we have to look
at the phase portrait of the clapper or at least value of the clapper
angular velocity just before the impact $\dot{\varphi}_{2,BI}$. We
focus on the clapper's motion because the changes in the bell's behavior
are often imperceptible due to its much bigger mass and inertia.

The two most common ringing schemes are symmetric ``flying'' and ``falling'' clapper. Analysing subplot (c) of Fig. \ref{fig:Lr-1a} we see that in ``flying clapper'' working scheme collisions have gentle course and the risk of damage of the bell or the clapper is minimal. In this regime the amount of energy dissipated during collision is relatively small and do not depend on the yoke design. Conversely in ``falling clapper'' ringing scheme the amount of energy dissipated at instant of collision strongly depends on the yoke design. Collisions are more abrupt but still the bell can operate reliably if the $l_r$ and $T$ parameters are properly chosen. The risk of failure is also higher for asymmetric attractors (both ``falling'' and ``flying'' clapper) because then the amount of energy transferred during collisions is significantly higher than in symmetric case (for similar parameters' values). That is why these working regimes are rarely met.

Vertical lines in Fig. \ref{fig:Lr-1a} mark the ranges of $l_{r}$
where given working regimes can be observed. We see that transitions between different states of the system are sudden 
and sometimes even minor changes in the yoke geometry can result in a completely different
behavior of the system. We usually expect the bells to work reliably
for decades without any maintenance. Therefore, the yoke and the propulsion
of the bell should be designed to ensure that the instrument will
work properly regardless of small changes of influencing parameters.
Such approach requires the in-depth knowledge of the systems dynamics.

\subsection{Two parameters ringing scheme diagrams}

To examine how the yoke's design - described by parameter $l_{r}$
- and the amplitude of forcing - parameter $T$ - influence the system's
dynamics we perform the series of numerical simulations. In our analysis
we consider the following ranges of these parameters: $l_{r}\epsilon\left\langle -1.3,\,0.23\right\rangle \,[m]$
and $T\epsilon\left\langle 100,\,625\right\rangle \,[Nm]$. We take
$154$ equally spaced values of $l_{r}$ (with the step $0.01\,[m]$)
and combine each with $42$ values of $T$ (from $100\,[Nm]$ to $625\,[Nm]$
with the step $12.5[Nm]$). That gives us $6468$ different sets of
system's parameters. For each one we simulate the system's motion
starting from zero initial conditions ($\varphi_{1}=0$, $\varphi_{2}=0$,
$\dot{\varphi}_{1}=0$, $\dot{\varphi}_{2}=0$). 

In Figs. \ref{fig:wykres1} and \ref{fig:wykres38} we show the possible
behavior of the systems $2$ minute and $38$ minutes after the excitation
started respectively. We call these two parameters plots as the ``ringing
scheme diagrams'' because they allow to determine which ringing scheme
can be achieved for the given values of $T$ and $l_{r}$. We consider
$7$ most characteristic types of the bells behavior that are described
in detail in the previous Section. Of course many different stable
attractors can exist for a given set of parameters $l_{r}$ and $T$,
but during our investigation we concern only the solutions that basins
of attraction contain zero initial conditions ($\varphi_{1}=0$, $\varphi_{2}=0$,
$\dot{\varphi}_{1}=0$, $\dot{\varphi}_{2}=0$). As we mention before,
we do so because in most cases the bell and the clapper start their
motion from hanging down position with zero velocities. Another aspect
that we consider is the transient time needed to reach a given attractor.
It is important in real applications because the launching procedure
of the bell should be possibly short. Hence, when designing the yoke
and the propulsion mechanism we have to know both how presumed ringing
scheme can be achieved as well as the time of transient behavior before
we reach that attractor. If this time is too long we can either change
the yoke design or apply special starting procedure - by the control
of the driving motor.

\begin{center}
\begin{figure}[H]
\begin{centering}
\includegraphics{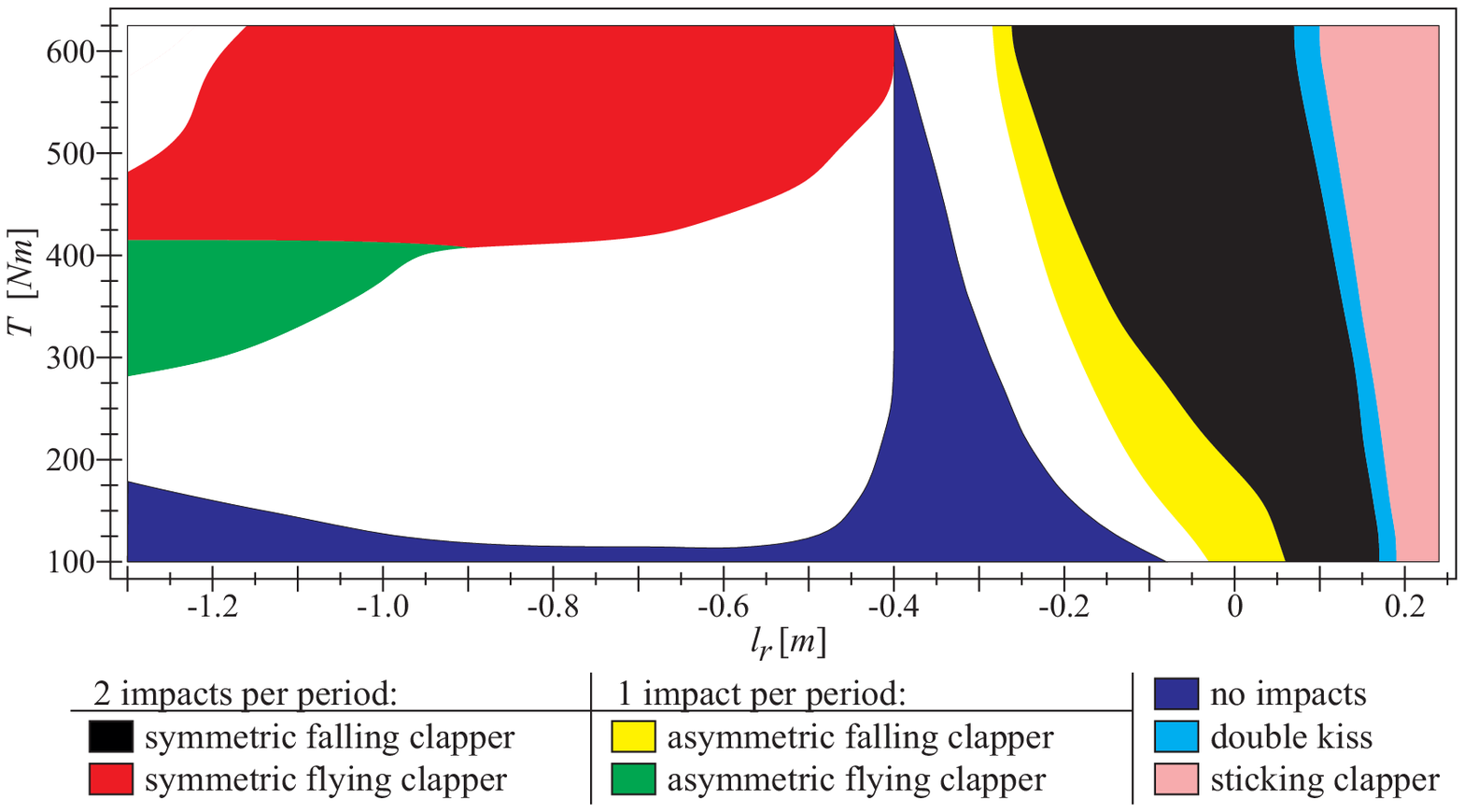}
\par\end{centering}

\protect\caption{Two parameter ringing schemes diagram showing the behavior of the
system $38$ minutes after the excitation started. \label{fig:wykres38} }
\end{figure}

\par\end{center}

The first plot - Fig. \ref{fig:wykres38} - presents the state of
the system $38$ minutes after the propulsion starts. It shows which
stable attractor will be achieved starting from zero initial conditions
for different values of $l_{r}$ and $T$. Hence, Fig. \ref{fig:wykres38}
indicates which ringing scheme can be obtained with given type of
yoke and maximal driving torque. Analyzing the plot we can draw the
following conclusions. The biggest areas correspond to the three most
common ringing schemes: \textquotedblleft falling clapper\textquotedblright{}
(black) and \textquotedblleft flying clapper\textquotedblright{} (red)
with 2 impacts per one period of motion and \textquotedblleft sticking
clapper\textquotedblright{} (pink) that is typical for the bells mounted
in the English manner. This means that these behaviors are relatively
easy to achieve and remain even after long period of time when values
of some parameters can change a bit (for example maximum torque generated
by the linear motor).

Generally, the design of the yoke determines how the bell will operate.
For $l_{r}<-0.4\,[m]$ we can observe ``flying clapper'' with $2$
(red) or $1$ (green) impacts per period but these ringing schemes
can be achieved only when $T$ is bigger than some threshold. The
threshold value decreases with the decrease of $l_{r}$ that causes
the growth of the inertia of the yoke-bell assembly with respect to
its axis of rotation. If $T$ is not sufficient we will not observe
any impacts (blue) or the system will reach different attractor (white)
- periodic or non-periodic one. For $l_{r}\epsilon\left(-0.4.\,-0.284\right)\,[m]$
we cannot obtain any of the analyzed ringing schemes no matter how
big the forcing amplitude is. This proves that when the yoke is designed
improperly it may be impossible to force the system to ensure proper
operation. For $l_{r}>-0.284\,[m]$ the system can work in the following
manner: ``falling clapper'' with $1$ (yellow) or $2$ (black) impacts
per period, ``double kiss'' (light blue) or ``sticking clapper''
(pink). Each of these ringing schemes can be achieved despite the
value of $T$ but the yoke should be designed for the purpose. Hence,
for these working regimes there is no need to use very powerful driving
motors and the system can work more efficiently. 

\begin{center}
\begin{figure}[H]
\begin{centering}
\includegraphics{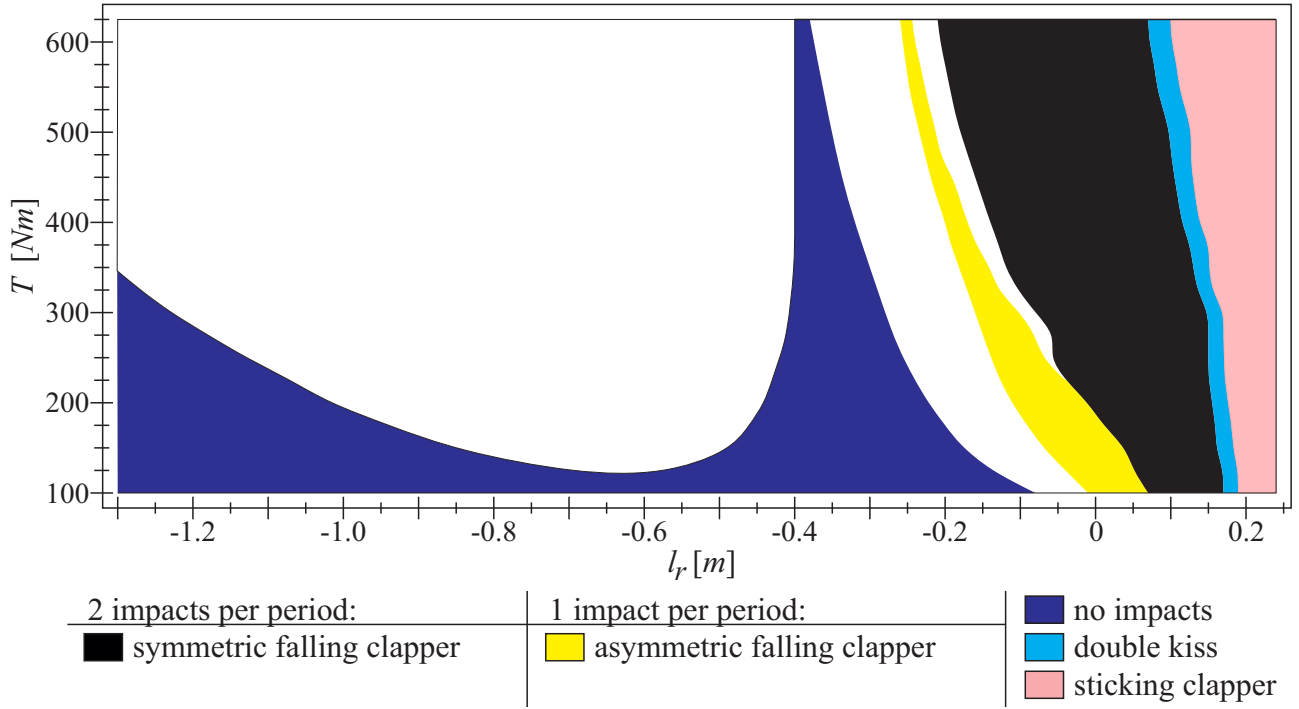}
\par\end{centering}

\protect\caption{Two parameter ringing schemes diagram showing the behavior of the
system $2$ minutes after the excitation started. \label{fig:wykres1}}
\end{figure}

\par\end{center}

In real applications we usually cannot accept launching times bigger
than two minutes. To determine for which parameters the start up procedure
runs quickly, we calculate the second plot - Fig. \ref{fig:wykres1}
- that shows the state of the system two minutes after the propulsion
started. In this case we analyze the system's motion for$120\,[s]$
and do not demand to reach the stable solution. Instead, we check
if for the average observer the behavior of the system could be seen
as proper and regular. For example we do not care if the amplitude
of the bell's motion is still slightly rising and just check if the
period between the successive impacts is constant and the collisions
have recurrent course. Comparing Figs. \ref{fig:wykres38} and \ref{fig:wykres1}
we see that the difference in the size of areas is especially visible
for $l_{r}<-0.4\,[m]$. The range that corresponds to the attractor
without impacts (blue color) is significantly larger and we cannot
observe any ``flying clapper'' ringing schemes. This means that
these working regimes not only requires larger amplitude of forcing
but also longer transient time before the system reaches presumed
solution. Therefore, in practical applications we often apply special
launching procedure to decrease the time of transient behavior. We
also see that for $l_{r}>-0.4\,[m]$ the areas corresponding to different
ringing schemes are almost identical to the ones presented in Fig.
\ref{fig:wykres38}. The main difference is that for $T>230\,[Nm]$
there is a white region between the areas corresponding to asymmetric
(yellow) and symmetric (black) ``falling clapper''. Therefore, special
starting techniques are rarely required if the yoke design ensures
small inertia with respect to the bell's rotational axis. 

Another important conclusion from the analysis of the ringing scheme
diagrams is the possibility to design the yoke and the propulsion
to enable more than one ringing scheme. For example, assuming that
$l_{r}=0$ we will observe ``falling clapper'' with $1$ impact
per period for $T\epsilon\left(100,\,191\right)\,[Nm]$ and $2$ impacts
per period for $T\epsilon\left(191,\,625\right)\,[Nm]$. Hence, if
the adopted driving motor enables to adjust the output torque in a
sufficiently wide range, we can change the way the bell operates.
Thanks to that, we are able to differentiate the tune of the bell
for different occasions.

\subsection{Example of the launching procedure}

The results presented in the previous Subsection prove that we often
have to use special launching procedure in order to achieve presumed
ringing schemes and maintain considerably short starting time. In
this Section we propose a simple control of the driving motor that
enables to achieve all considered working regimes after only two minutes.
We assume that launching procedure takes $120$ seconds and over that
time we adjust the maximum torque that is generated by the motor.
Hence, during the first $120$ seconds, instead of using formula \ref{eq:Mt}
to describe the generalized momentum generated by the propulsion mechanism,
we use the following function:

\begin{equation}
M_{t}(\varphi_{1},t)=\begin{cases}
\begin{array}{ccc}
T\left(1+\left(\frac{T\,|l_{r}|}{D_{1}}-\frac{T\,|l_{r}|}{D_{2}}t\right)^{2}\right)\textrm{sgn}(\dot{\varphi}_{1})\,\cos\left(7.5\varphi_{1}\right), & \: if & |\varphi_{1}|\leq\frac{\pi}{15}\\
\\
0, & \: if & |\varphi_{1}|>\frac{\pi}{15}
\end{array}\end{cases}\label{eq:Mt-control}
\end{equation}

where: $t$ is time and $D_{1}=200\,[Nm^{2}]$, $D_{2}=24000\,[Nm^{2}s]$
are control parameters. We obtained the above formula basing on the
following assumptions. The amplitude of forcing decreases as a square
function and for $t=120\,[s]$ it is equal to parameter $T$. Hence,
there is no jump in the torque generated by the motor when the launching
procedure ends ($t=120\,[s]$) and we switch back to formula \ref{eq:Mt}.
Analyzing Figs. \ref{fig:wykres38} and \ref{fig:wykres1} we see
that the launching procedure is needed especially for small $l_{r}$
and large $T$. Therefore, we assume that the maximal value of generated
torque should increase with the increase of $T$ and the absolute
value of $l_{r}$.

\begin{center}
\begin{figure}[H]
\begin{centering}
\includegraphics{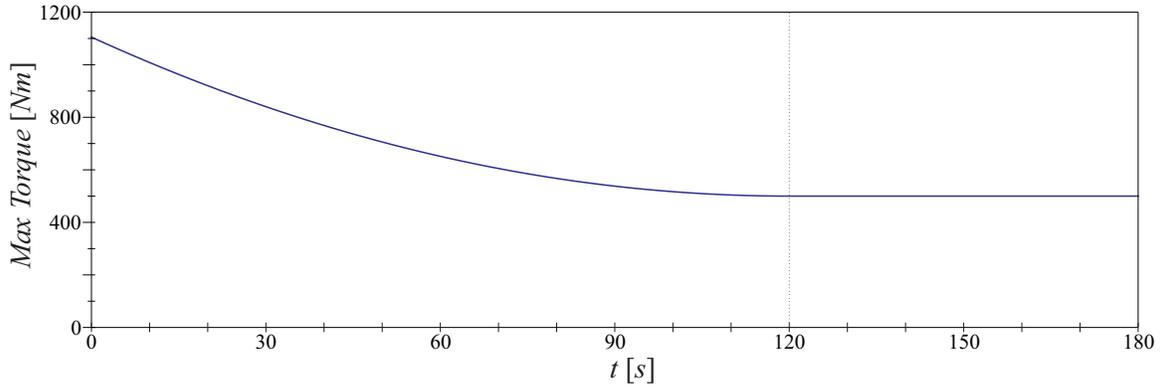}
\par\end{centering}

\protect\caption{Maximum torque generated by the linear motor during the first $3$
minutes of motion. \label{fig:Max_torque}}
\end{figure}

\par\end{center}

In Fig. \ref{fig:Max_torque} we show how the maximum torque generated
by the linear motor (for $T=500\,[Nm]$ and $l_{r}=-0.44\,[m]$) changes
during the analyzed first $3$ minutes of the system's motion. The
dotted vertical line in Fig. \ref{fig:Max_torque} marks the time
when the launching procedure ends and the output of the motor stabilizes. 

\begin{center}
\begin{figure}[H]
\begin{centering}
\includegraphics{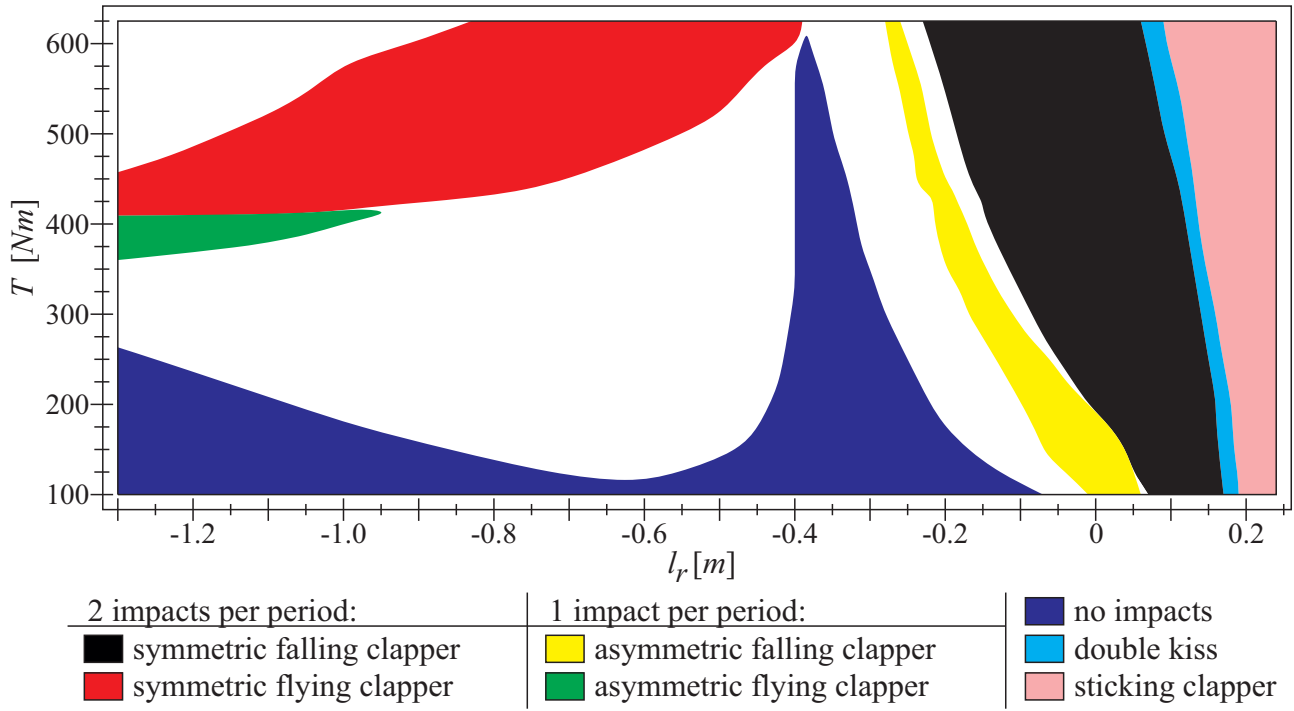}
\par\end{centering}

\protect\caption{Two parameter ringing schemes diagram showing the behavior of the
system with applied launching procedure ($2$ minutes after the excitation
started). \label{fig:wykres_control}}
\end{figure}

\par\end{center}

In Fig. \ref{fig:wykres_control} we present the ringing scheme diagram
obtained for the system with applied launching procedure that is described
above. Similarly to Fig. \ref{fig:wykres1}, this time we also do
not require the system to reach the stable attractor and just check
if its behavior could be perceived as proper by an average listener.
Analyzing Fig. \ref{fig:wykres_control} we see that even using simple
control of the driving motor we can achieve satisfactory results and
decrease the time of the transient motion. Moreover, obtained results
prove that formula \ref{eq:Mt-control} - universal for all $T$ and
$l_{r}$ - allow us to achieve good results in a wide range of parameters
values. But, the proposed control is not efficient with respect to
the systems with large forcing amplitude and yoke design that provides
large inertia of the bell (left upper part of the plot.). Comparing
Figs. \ref{fig:wykres38} and \ref{fig:wykres_control} we see the
difference in the borders and sizes of the areas corresponding to
different working regimes, but still thanks to the proposed launching
procedure we can reach all $7$ considered working regimes. Hence,
the proposed control allow us to obtain all considered ringing schemes
but they appear in smaller ranges of parameters values. Therefore,
in some cases there is a need to adjust the launching procedure to
shorten transient time and reach presumed behavior of the system.

\section{Conclusions}

Hybrid dynamical model of the yoke-bell-clapper system can exhibit
a plethora of different dynamical behaviors but only a few of them
can be considered as a proper working regimes of the instrument. Therefore,
the yoke of the bell and its propulsion mechanism should be designed
to ensure the system will behave correctly and sound nicely. In case
of bells with typical shape and mass engineers usually design yokes
basing on their experience and local tradition. Still, it is often
a challenge to create a reliable suspension for unique bells and sometimes
the mounting of the bell or its propulsion have to be redesigned to
achieve presumed ringing scheme.

In the paper we present a method that can be used to determine the
conditions under which given type of behavior can be achieved. We
develop two parameter ringing scheme diagrams that describe how the
geometry of the yoke and maximum output of the driving motor influence
the dynamics of the system. Similar charts can be calculated for any
bell and used to design its mounting and propulsion. Thus, we can
ensure that the instrument will work properly and reliably regardless
of small changes of parameters. 

Moreover, analyzing the ringing scheme diagrams we see that it is
possible to design the yoke and the propulsion to enable more than
one ringing scheme with relatively small changes of motor and/or yoke
parameters. Hence, we can easily differentiate the tune of the bell
for different occasions.

Basing on the presented analysis we are able to determine the minimum
required power of the driving motor and estimate the time that is
needed to achieve given attractor. If the transient time is too long,
special launching procedure has to be applied. We propose the control
of the driving motor that enables to reach all considered ringing
schemes after $120$ seconds of preliminary motion. It is possible
to further optimize the launching procedure, but to get best results
we need to know the values of all system's parameters. Still, presented
results are robust and prove that even simple and universal formula
can be highly efficient in a wide range of systems' parameters.

\section*{Acknowledgment}

This work is funded by the National Science Center Poland based on
the decision number DEC-2013/09/N/ST8/04343. We would especially like
to thank the Parson of Cathedral Basilica of St Stanislaus Kostka
Prelate Ireneusz Kulesza for his support and unlimited access to the
bell. We have been able to measure the bell's template thanks to the
bell's founder Mr Zbigniew L. Felczynski. The
data on the clapper, the yoke and the motor have been obtained form
Mr. Pawel Szydlak.

\section*{References}

\bibliographystyle{plainnat}

\end{document}